\documentstyle[12pt,psfig,cite,a4]{article}
\begin{document}
\begin{titlepage}
\begin{flushright}
KEK-TH-685
\end{flushright}
\vspace{0.5cm}
\Large

\begin{center}
Phase Structure of \\
Four-dimensional Simplicial Quantum Gravity \\ 
with a $U(1)$ Gauge Field
\end{center}
\vspace{0.5cm}
\begin{center}
\large
S.Horata \footnote[1]{E-mail address: horata@ccthmail.kek.jp}, 
H.S.Egawa \footnote[2]{E-mail address: egawah@ccthmail.kek.jp}, 
N.Tsuda \footnote[3]{E-mail address: tsudan@ccthmail.kek.jp
} and 
T.Yukawa \footnote[4] {E-mail address: yukawa@post.kek.jp}
\vspace{0.3cm}

\normalsize
$^{\ast, \dag, \ddag, \S}$ 
Theory Division, Institute of Particle and Nuclear Studies, \\
KEK, High Energy Accelerator Research Organization, \\
Tsukuba, Ibaraki 305-0801, Japan \\ 

\vspace{0.3cm}
$^{\dag}$ 
Department of Physics, Tokai University, \\
Hiratsuka, Kanagawa 259-1292, Japan \\
%

%
%

\vspace{0.3cm}
$^{\S}$ 
Coordination Center for Research and Education, \\ 
The Graduate University for Advanced Studies, \\ 
Hayama-cho, Miuragun, Kanagawa 240-0193, Japan 

\end{center}

\vspace{0.5cm}
\begin{abstract}
The phase structure of four-dimensional simplicial quantum 
gravity coupled to $U(1)$ gauge fields has been studied using 
Monte-Carlo simulations. 
The smooth phase is found in the intermediate region between 
the crumpled phase and the branched polymer phase. This new 
phase has a negative string susceptibility exponent, even if 
the number of vector fields ($N_{V}$) is $1$. 
The phase transition between the crumpled phase and the smooth 
phase has been studied by a finite size scaling method. 
From the numerical results, we expect that this model (coupled 
to one gauge field) has a higher order phase transition than first 
order, which means the possibility to take the continuum limit 
at the critical point. 
Furthermore, we consider a modification of the balls-in-boxes 
model for a clear understanding of the relation between the numerical 
results and the analytical one. 
\end{abstract}

\end{titlepage}
\section{Introduction} 
To formulate a theory of quantum gravity in 
four dimensions, many approaches have been tried. 
%
One of them is a numerical approach with the method of 
dynamical triangulation. 
In two dimensions, the quantum gravity can be quantized 
for a central charge $c<1$. 
The method of dynamical triangulation has generally been 
considered to be a correct discretized model, and has given 
consistent results with the analytical approach: for example, 
the MINBU analysis\cite{JM,AJT} and the loop length 
distributions\cite{TY,KKMW}.
Two-dimensional quantum gravity has generally been regarded 
as being a toy model of four-dimensional quantum gravity. 
Recently, a numerical approach with the method of dynamical 
triangulation in analogy with a two-dimensional model has been 
studied. 
In the four-dimensional case, it has been expected that the phase 
transition point between the strong coupling phase and the weak 
coupling phase is statistically the second phase transition point. 
Moreover, this point is recognized as the ultraviolet (UV) fixed 
point of the quantum theory of gravity.

From numerical results, in four-dimensional pure gravity, 
it is known that there are two distinct 
phases\cite{AJ,AM,CKR,HIN,CKRT,EHITY}.
For small values of the bare gravitational coupling constant, 
the system is in the so-called elongated phase, which has the 
characteristics of a branched polymer phase. 
For large values of the bare gravitational coupling constant, 
it is in the so-called crumpled phase.
Numerically, the phase transition between the two phases has 
actually been shown to be of first order\cite{BBKP,BAKKER}.
Therefore, it is difficult to construct a continuum theory. 
In other words, It is difficult to define the quantum theory 
of gravity on a four-dimensional triangle (4-simplex) 
as a simple application of the two-dimensional lattice model.

Our next step is to investigate the possibilities to extend 
the model of four-dimensional quantum gravity.
We have three motivations: (1) a modified model\cite{HIN} 
in three-dimensional case, which suppresses the vertex order 
concentration (VOC) as the singular sub-simplex, and which can be 
changed of the phase structures, (2) the property of quantum 
field theory with background metric independence\cite{AMM,HS,HAMA}, 
that the manifold may be made stable by adding matter
fields and (3) the balls-in-boxes model, that gives a scenario for a 
phase transition in four-dimensional simplicial quantum
gravity\cite{BBPT,BBJ}, and that shows possibilities to change
the phase structure with some modifications of the model\cite{BB}.
For the possibility of a continuum theory, at least, we have to
find a new phase structure that has a second order phase 
transition point. 
One modification is to introduce gauge matter fields\cite{BILKE}.

Recently, the phase structure with vector fields has been studied
numerically\cite{BILKE2,PHASE-DIAGRAM}.
In the case of a model with vector fields, the phase structure is
changed drastically and the intermediate phase, the so-called smooth
phase, has been observed between the crumpled phase and the elongated
phase. 
In this region, the string susceptibility exponents ($\gamma_{st}$) 
have negative values. 
They show that this region has a fractal property, and may be smooth 
compared with the branched polymer region.
We thus expect the possibility of a continuum limit at the critical 
point between the crumpled phase and the smooth phase.
In order to investigate the nature of the phase transition, we measure 
the critical exponent as the finite size scaling, and also study 
the scaling property of the mother boundary in analogy with 
the the two-dimensional case, and we expect that the scaling structure 
also appears in the boundary in the four-dimensional case. 

This paper is organized as follows. In Section 2, we discuss the model
of four-dimensional dynamical triangulation with some vector fields.
In Section 3, we show our numerical results concerning measurements of 
the string susceptibility exponents ($\gamma_{st}$) and a schematic phase
diagram. We thus discuss the statistical property of the phase 
transition between the crumpled phase and the smooth phase in the case
of four-dimensional simplicial quantum gravity coupled to one gauge
field ($N_{V}=1$). 
Furthermore, in section 4, we discuss the scaling property near to 
the critical point. In section 5, we discuss a scenario for 
the phase structure and the phase transition in four-dimensional 
simplicial quantum gravity coupled to matter fields. Finally, we 
discuss the possibility of a continuum limit of four-dimensional 
simplicial quantum gravity in this article. 
%
\section{Model}
It is still not known how to provide a constructive definition 
of four-dimensional quantum gravity.
We have considered a discretized random closed manifold in 
analogy with the two-dimensional case.
We numerically evaluated the Euclidean path integral with the technique
of dynamical triangulation, which gives a discrete summing over all 
possible connections of lattices that may replace the integral over
diffeomorphism inequivalent metrics.
Then, we naturally considered the Euclidean Einstein-Hilbert action 
coupled to $N_V$ copies of $U(1)$ vector fields and its discretized 
model with 4-simplices. 
The total action is $S = S_{EH} + S_{pl}$. 
We use the Einstein-Hilbert term for gravity, 
\begin{equation}
S_{EH}[\Lambda,G] 
= \displaystyle{\int} d^{4}x \sqrt{g}\left(\Lambda-\frac{1}{G}R \right), 
\label{eq:1}
\end{equation}
where $\Lambda$ is the cosmological constant and $G$ is Newton's
constant. 
We use the discretized action for gravity, 
\begin{eqnarray}
 S_{EH}[\kappa_{2},\kappa_{4}] 
&=& \kappa_{4}N_{4}-\kappa_{2}N_{2} \nonumber \\
&=& - \frac{2 \pi}{G} N_2 + \left( \Lambda' + \frac{10}{G} 
\cos^{-1} \left( \frac{1}{4} \right) \right),
\label{eq:2}
\end{eqnarray}
where $\kappa_{2} \sim 1/G$, $\kappa_{4}$ is related to $\Lambda'=c
\Lambda$ (c is the unit volume) and $N_{i}$ is the number of
$i$-simplices. 
We use the plaquette action for $U(1)$ gauge fields\cite{BILKE}, 
\begin{equation}
S_{pl}=\sum_{t_{ijk}}o(t_{ijk})[A(l_{ij})+A(l_{jk})+A(l_{ki})]^{2},
\label{eq:3}
\end{equation}
where $l_{ij}$ denotes a link between vertices $i$ and $j$, $t_{ijk}$ denotes 
a triangle with vertices $i$, $j$ and $k$ and  $o(t_{ijk})$ denotes the
number of 4-simplices sharing triangle $t_{ijk}$. We consider the $U(1)$ gauge
field $A(l_{ij}) = - A(l_{ji})$ on a link $l_{ij}$.

We consider that a partition function of gravity with $N_V$ copies of
$U(1)$ gauge fields is
\begin{equation}
Z(\kappa_{2},\kappa_{4},N_V) = \sum_{N_4} e^{-\kappa_4 N_4} 
\sum_{t(2D) \in T(4D)} e^{\kappa_2 N_2} \prod_{N_V} 
\int \prod_{l \in t(2D)} dA(l) e^{-S_{pl}}.
\label{eq:4}
\end{equation}
We sum over all four-dimensional simplicial triangulation 
$T(4D)$ in order to carry out a path integral over the metric, 
where we fix the topology with $S^{4}$.
As is well known, we must add a small correction term 
($\delta S$)\cite{AM} to the lattice for fluctuations of volume, 
\begin{equation}
 \delta S = \delta \kappa_2 (N_4 - N_4^{target})^2, 
\label{eq:5} 
\end{equation} 
where $\delta \kappa_2$ is adjusted with an appropriate choice; 
we use $\delta \kappa_2 = 0.00025$. 

Near to the critical point, it is expected that the partition function
(eq.(\ref{eq:4})) behaves as 
\begin{eqnarray}
Z(k_2 \sim k_2^c ) 
&\sim& \sum_{T(4D)} N_4^{\gamma_{st}-3} e^{\mu N_2} \nonumber \\ 
&\sim& 1/|N_4 - N_4^{target}|^{3-\gamma_{st}}, 
\label{eq:6}
\end{eqnarray}
where $\gamma_{st}$ is the string susceptibility exponent 
related to the entropy of the manifold. In the situation; 
$\gamma_{st} > 0$, the manifold grows to the spiky configuration. 
Conversely, in $\gamma_{st} < 0$, the surfaces grow to smooth 
structures. 

Using Monte-Carlo simulations, we evaluated the partition function 
(eq.(\ref{eq:4})). We followed the way of updating the configuration
after ref.\cite{BILKE}. This is that the gauge fields renew the sequence 
of the $(p,q)$-move according to the weight of the Bolzman factor. 
The action was checked by the Metro police methods. As we held the same 
conditions with Bilke et al., we performed a geometry update, 
$(p,q)$-move, updating the gauge fields by the heat-bath sweep and 
the over-relaxation sweep. An over-relaxation sweep was introduced 
for the convergence, 
\begin{equation}
 A_{ij} \rightarrow A_{ij} - 2 \bar{A}_{ij}.
  \label{eq:7}
\end{equation}
The measurement chance $(N_4=N_4^{target})$ came at intervals of 
about 100 sweeps, where we counted ``1 sweep'' as the flow of 
``heat-bath sweep $\rightarrow$ $(p,q)$-move $\rightarrow$ 
over-relaxation sweep''. 
The measurement processes and updating processes are the same as the pure 
gravity case. However, since the case of adding vector fields costs more CPU 
time, we arranged for suitable compute time. 
%
\section{Phase diagram with vector fields}
%
In order to investigate the phase structure of simplicial quantum 
gravity, first we calculated the string susceptibility exponent 
($\gamma_{st}$). This exponent is defined by the asymptotic form of 
the partition function (eq.(\ref{eq:6})). 
It is known that the case of $\gamma_{st}$ corresponds to the
dominance of the branched polymer\cite{KAWAI}. 
We measure the string susceptibility exponent ($\gamma_{st}$) 
with the method of the MINBU (Minimum Necked Baby Universe)
analysis\cite{JM,AJT,COMMON}. It is a powerful exponent for probing
the property of quantum geometry, and is easily measured. We can count 
a baby universe which is connected to the mother universe via a minimal
neck. The distribution function for the MINBU analysis with the size B 
can written as 
\begin{eqnarray}
 n_A &=& \frac{3(A-B+1)(B+1)Z[A-B+1]Z[B+1]}{Z[A]} \nonumber \\
     &\sim& c \left\{ \Bigl( B+1 \Biggl) \Biggl( 1 - \frac{B-1}{A} \Biggl) 
     \right\}^{\gamma_{st}-2},
\label{eq:8}
\end{eqnarray}
where we use the asymptotic form of the partition function, 
$Z[A] \sim A^{\gamma_{st}-3} e^{\mu A}$, where $A$ denotes 
the volume of the mother universe. 
%

In the pure gravity case, a measurement of the string susceptibility 
exponent is discussed in ref.\cite{COMMON}. 
From the numerical results, the string susceptibility exponent 
supports the idea that the pure case of simplicial quantum gravity has 
two distinct phases. 
%
%
%
What is important in the $N_{V}=1$ case is that the usual phase 
transition point ($\kappa_{2}^{c}$) is different from another 
transition point ($\kappa_{2}^{o}$), which separates 
the $\gamma_{st} < 0$ region from the $\gamma_{st} > 0$ region, 
and $\gamma_{st}$ becomes negative on the phase transition point 
($\kappa_{2}^{c}$). 
This fact leads to the definition of a new smooth phase. 
This phase is defined by an intermediate region between these 
two transition points ($\kappa_{2}^{c}$ and $\kappa_{2}^{o}$). 
In the pure gravity case, it is clear that 
$\kappa_{2}^{c} \approx \kappa_{2}^{o}$, 
and thus there is no evidence for the existence of a new smooth phase. 
On the other hand, in the case of $N_{V}=1$ with $N_{4}=16K$, we observe 
the $\gamma_{st} < 0$ region beyond the usual phase transition point 
($\kappa_{2}^{c}$). 
We also observe a very obscure transition from $\gamma_{st} < 0$ to 
$\gamma_{st} > 0$ at $\kappa_{2}^{o}$. 

\begin{figure}
\centerline{\psfig{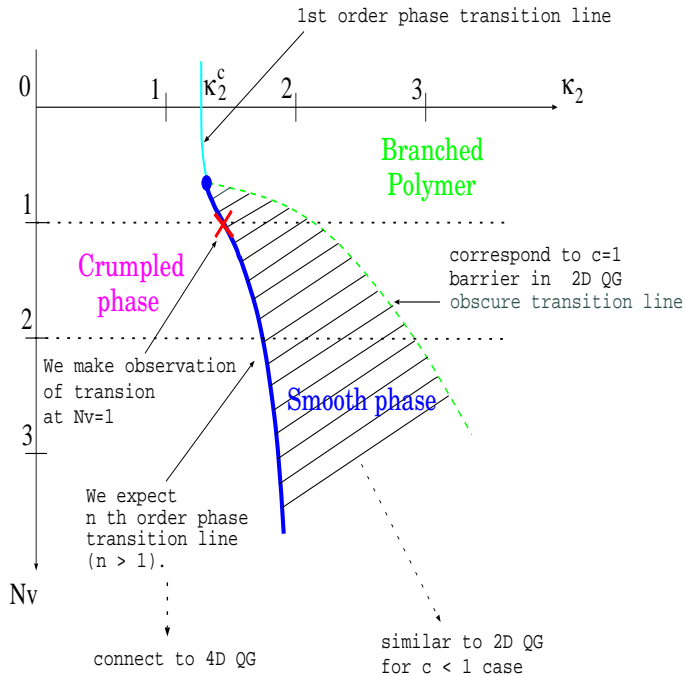}}
\caption
{
Schematic phase diagram of four-dimensional simplicial quantum gravity. 
}
\label{fig:phasediagram} 
\end{figure}

We give the numerical results at Table.\ref{table:2}.
\begin{table}[hbt]
 \caption{Measurement results of $\gamma_{st}$ at $N_4 = 16K$.}
 \label{table:2}
 \begin{center}
  \begin{tabular}{c|c|c|c}
   \noalign{\hrule height0.8pt}
   $\kappa_2$ & $N_V = 0$ & $ N_V = 1$ & $N_V = 2$ \\ \hline
   1.0        & -0.43(5)  & -2.4(3)    & -3.9(3)   \\
   1.5        &  0.46(2)  & -0.87(9)   & -2.4(2)   \\
   2.0        &  0.53(1)  & -0.07(3)   & -1.05(9)  \\
   2.5        &  0.59(2)  &  0.136(4)  & -0.45(5)  \\
   3.0        &  0.67(1)  &  0.123(3)  & 0.01(4)   \\
   3.5        &    -      &  0.188(7)  & 0.15(8)   \\
   \noalign{\hrule height0.8pt}
  \end{tabular}
 \end{center}
\end{table}
%
\begin{figure}
\centerline{\psfig{file=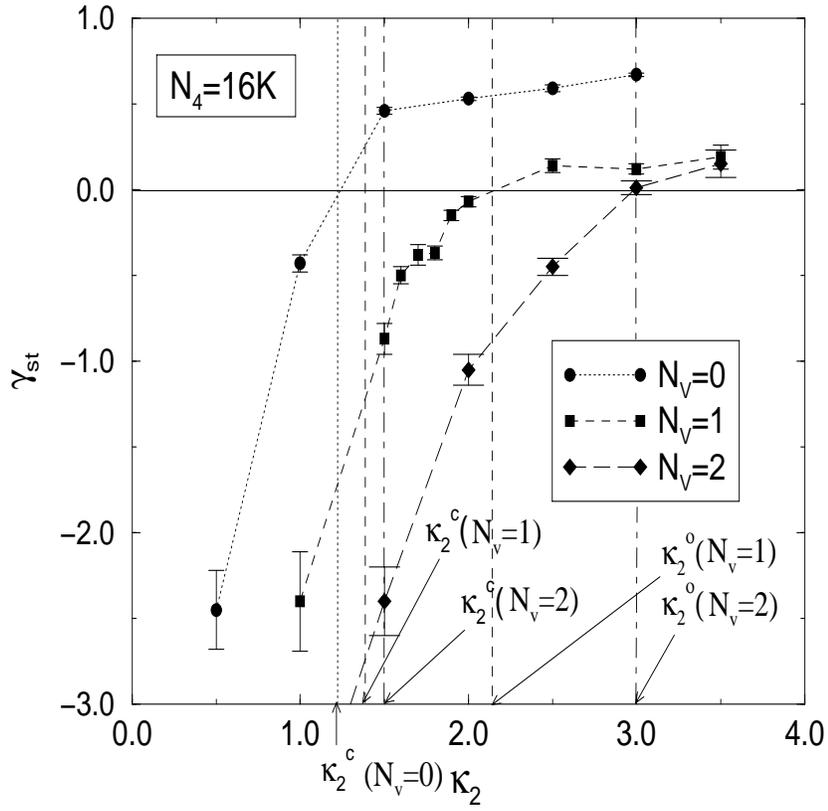,height=11cm,width=11cm}}
\caption
{
String susceptibility exponents ($\gamma_{st}$) plotted 
versus the coupling constants ($\kappa_{2}$) for $N_{V}=0$, 
$1$ and $2$. 
}
\label{fig:gamma}
\end{figure}
%
In Fig.\ref{fig:phasediagram}, we plot $\gamma_{st}$ for various 
numbers of gauge fields versus $\kappa_{2}$ with volume $N_{4}=16K$.
In the case of adding vector fields, we can find that the usual
phase-transition point ($\kappa_{2}^{c}$) is different from another
transition point ($\kappa_{2}^{o}$) which separates the $\gamma_{st} <
0$ region from the $\gamma_{st} > 0$ region and $\gamma_{st}$ becomes
negative at the phase-transition point ($\kappa_{2}^{c}$). 
This fact leads to the existence of a new phase. We thus call this 
intermediate region the smooth phase between these two transition
points ($\kappa_{2}^{c}$ and $\kappa_{2}^{o}$). 
%
%

We can consider that simplicial quantum gravity coupled to 
vector fields has three phases\cite{BILKE2,PHASE-DIAGRAM,EHTY}. 
In Fig.\ref{fig:phasediagram}, we show a schematic phase diagram.
%
%
We have three phases in this parameter space: a crumpled phase, 
a smooth phase (shaded portion) and a branched polymer phase. 
Furthermore, the smooth phase expands with adding more vector fields. 
We show the discontinuous phase transition as the thin line and 
the smooth phase transition as the thick line. 
In the case of adding the vector fields, there are two separate phase
transition lines: the usual phase transition line ($\kappa_2^{c}$) and
an obscure phase transition line ($\kappa_2^{o}$). 
The obscure transition at $\kappa_{2}^{o}$ has been shown
to be third order or cross over\cite{BBPT}, which is very similar to 
the $c=1$ barrier in two-dimensional quantum 
gravity\cite{PHASE-DIAGRAM}. 
In two dimensions the $c=1$ barrier is well-known as 
an obscure transition from the fractal phase 
($c \leq 1$ and $\gamma_{st}^{2D} < 0$) 
to the branched polymer phase 
($c > 1$ and $\gamma_{st}^{2D} > 0$).
We consider that the obscure phase transition point may be a 
threshold point, like the $c=1$ case in two dimensions. 
From reports of Antoniadis et al.\cite{AMM}, the quantum field
theory of gravity with conformal invariance has a central charge 
$Q^2$, which has a threshold value, $Q^2 \approx 8$. 
We expect that the obscure phase transition point that is defined by the
threshold value, $\gamma_{st} \approx 0$, may be the same situation as 
the $Q^2 \approx 8$ case. 
Another phase transition point is the usual phase transition point 
between the strong coupling region (the crumpled phase) and 
the intermediate region (the smooth phase). 
We expect that the phase transition at $\kappa_{2}^{c}$ is 
continuous. This leads to the continuum limit of four-dimensional 
quantum gravity. 

\begin{figure}[t]
\centerline{\psfig{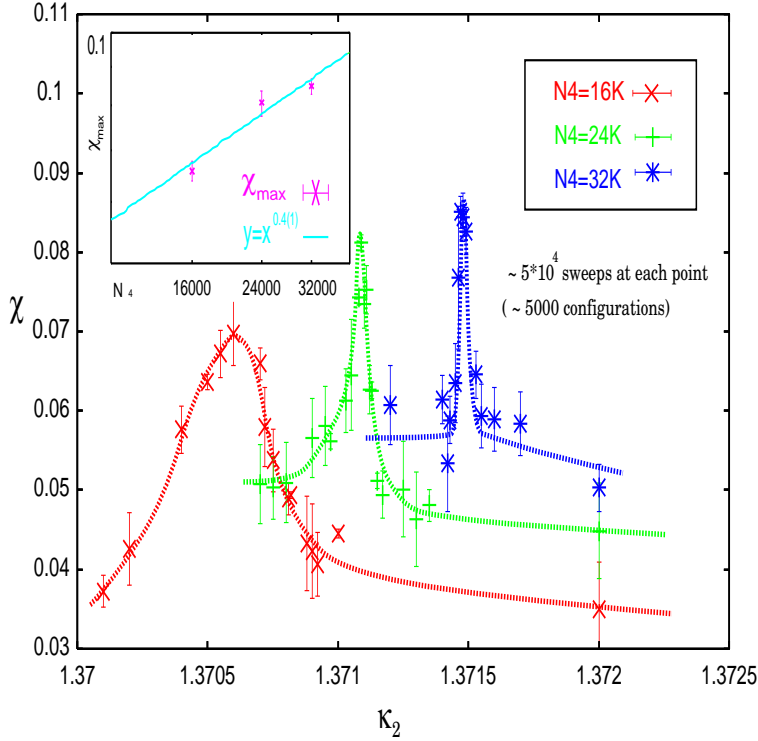}} 
\caption
{
Node susceptibility ($\chi$) plotted versus 
the coupling constant ($\kappa_{2}$) for $N_{V}=1$.
}
\label{fig:fss}
\end{figure}

Now, let us take a look at the transition at $\kappa_{2}^{c}$ in the
case of $N_V=1$.
In order to investigate the phase transition, we observe at the
exponents of the node susceptibility. The node susceptibility is 
defined in ref.\cite{BAKKER} as follows: 
\begin{equation}
 \chi = \frac{1}{N_4} \left( <N_0^2> - <N_0>^2 \right).
\label{eq:9}
\end{equation}
In Fig.\ref{fig:fss}, we plot the node susceptibility($\chi$) as a
function of $\kappa_2$ with the volume $N_{4}=16K, 24K$ and $32K$,
respectively. 
%
%
The node susceptibility ($\chi$) has a peak value 
at the critical point ($\kappa_2^c$). 
We find the peak value in each size. 
The height and the width of the susceptibility peaks give the finite
size scaling exponents of the phase transition.
The peak value ($\chi_{max}$) and the width of peak ($\delta \kappa_2$) 
grow as $N_{4}$ in power.
The susceptibility exponents ($\Delta$ and $\Gamma$) are defined by
\cite{BAKKER}: 
\begin{equation}
 \chi_{max} \propto N_{4}^{\Delta},
 \quad (\delta \kappa_2 \propto N_{4}^{-\Gamma}).
\label{eq:10}
\end{equation}
From the numerical results (Fig.\ref{fig:fss}), we obtain 
the susceptibility exponents: 
$\Delta = 0.4(1)$, $\Gamma \sim 0.5(3)$.
These values are apparently smaller than 1, 
though they are 1\cite{BAKKER} in the pure gravity case. 
This numerical results show that simplicial quantum gravity coupled to
a vector field has the different type of phase transition from simplicial 
quantum gravity in the pure gravity case. 

\begin{figure}
\centerline{\psfig{file=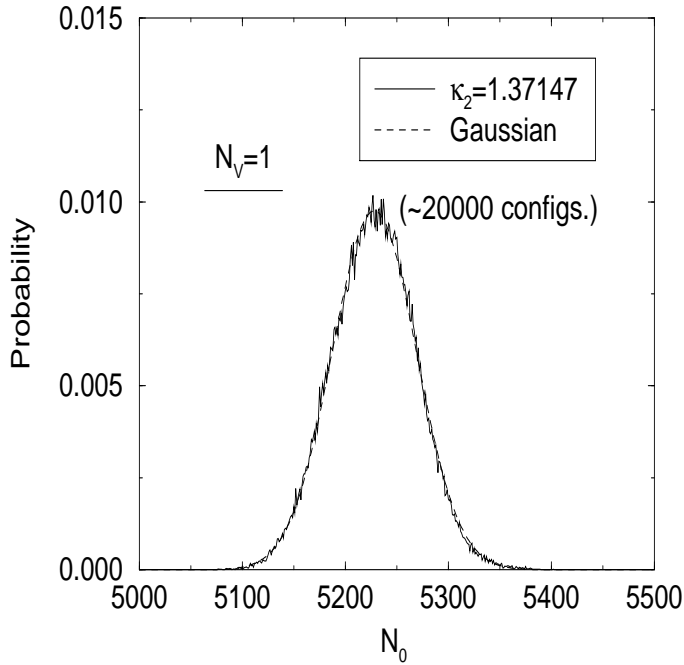,height=9cm,width=9cm}} 
\caption
{
Histogram of node ($N_0$) near to the critical point 
($\kappa_2^{c} = 1.37147(1)$) for $N_{V} = 1$ with $N_{4} = 32K$. 
The number of configurations is 10,000. 
}
\label{fig:node}
\end{figure}

In Fig.\ref{fig:node}, we show a histogram of $N_0$ for a size of 
$N_{4} = 32K$ near to the critical point 
($\kappa_2^{c} = 1.37147(1)$). 
%
%
In the pure gravity case, previously, a double peak structure 
has been found\cite{BBKP,BAKKER}. This structure relates to 
the latent energy. The fact that the phase transition is first order 
is shown. 
However, by adding the vector field, the double peak structure
disappears. 
We thus conclude that the phase transition between the crumpled 
phase and the smooth phase may be continuous, not first order. 
%
\section{
Scaling property of four-dimensional \\ 
simplicial quantum gravity
}
%
Recent numerical results obtained by dynamical triangulation in 
four space-time dimensions suggest the existence of the scaling behavior,
for example, the MINBU distribution\cite{COMMON}, correlation functions
\cite{BAKKER-SMIT} and the boundary volume distributions\cite{EHITY}.
If we assume the existence of a correct continuum limit, the
scaling property of quantum geometry gives important information
about the continuum theory, because some scaling properties are related to
the universality of the theory. 
One of the interesting observes is the fractal dimension 
(Hausdorff dimension) ($d_H$), 
\begin{equation} 
 d_H = \frac{d \ln V^{(4)}(D)}{d \ln D}, 
 \quad ( V^{(4)} \sim D^{d_H} ). 
\label{eq:11}
\end{equation}
This is based on studying the behavior of the volume $V^{(4)}(D)$, 
the number of simplices, within a geodesic distance $D$.
The geodesic distance $D$ is defined as the shortest distance between
two simplices through the center of the simplices.
In the pure gravity case, the Hausdorff dimension shows a different behavior 
in each phase. In the crumpled phase, the Hausdorff dimension diverges 
($d_H \rightarrow \infty$). 
The Hausdorff dimension rises very steeply to 
a large value below the transition, while it falls very rapidly to 
a branched polymer, $d_H = 2$, above the transition. 
However, in the case of gravity coupled to vector fields, each of three 
phases shows a different behavior for the Hausdorff dimension. 
In the crumpled phase and the branched polymer phase, the behavior for 
the Hausdorff dimension is similar to that of pure gravity. 
Furthermore, we find that the smooth phase is the intermediate region with
$2 < d_H \sim 4$.
For the smooth phase, the value of the Hausdorff dimension ($d_H$) is 
changed to smooth, as compared with the pure gravity case.
This fact supports the results of a finite size analysis.
Especially, at the critical point, we observe the Hausdorff dimension, 
$d_H = 4.6(2)$, with $N_4 = 32K$. 

Next, let us discuss the scaling structure of four-dimensional 
DT mfd, at the focusing on the scaling structure of the boundaries 
in four-dimensional Euclidean space-time using the concept 
of geodesic distances. 
We consider that the scaling structure of the boundaries has more 
informations than the Hausdorff dimension, and that it is the way 
of directly searching for the structure of quantum geometry. 
%
%
\begin{figure}
\centerline{\psfig{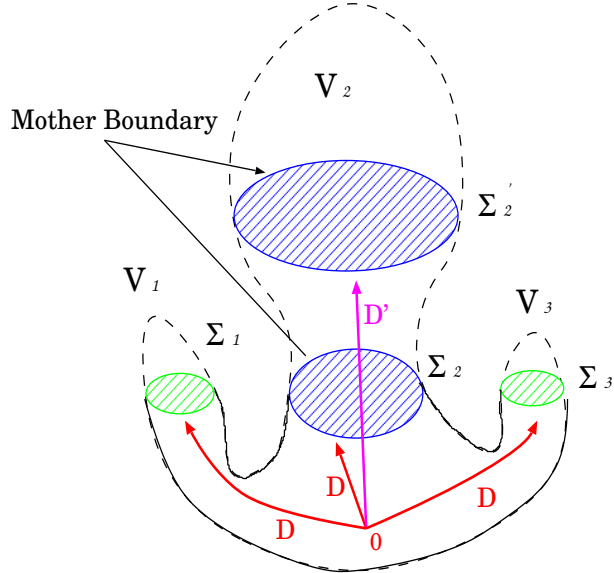}} 
\caption
{
Schematic picture of the boundaries ($\Sigma_{1},\Sigma_{2}$ and 
$\Sigma_{3}$) at distance $D$ and a boundary ($\Sigma_{2}'$) 
at distance $D'$ in four-dimensional Euclidean space with $S^{4}$ 
topology. The mother boundary is defined as a boundary ($\Sigma_{2}$) 
with the largest tip volume ($V_{2}$); the other boundaries 
are defined as baby boundaries. 
}
\label{fig:universe}
\end{figure} 
%
In the pure gravity case, the scaling property of the boundary volume 
is discussed in ref.\cite{EHITY}. 
As in the previous analysis\cite{EHITY}, we assume that 
the boundary volume distribution ($\rho(x,D)$) is a function of 
a scaling variable in analogy with the loop length distribution 
function in the two-dimensional model. 
Fortunately, in the two-dimensional model, the loop length distribution
function has been calculated analytically \cite{KKMW}.
The loop length distribution function, $\rho(x=L/D^2, D)$, which gives the
probability of the boundary loop with the loop length ($L$) within a
geodesic distance ($D$), is given as a function of the scaling
variable $x=L/D^2$: 
\begin{equation}
\rho(x=L/D^2, D) = \frac{3}{7\sqrt{\pi}} \frac{1}{D^2} \biggl(
x^{-3/2} + \frac{1}{2} x^{-5/2} +\frac{14}{3} x^{1/2} \biggl) e^{-x}.
\label{eq:12}
\end{equation}
This distribution function consists of two different types of 
distributions. 
The first two term of eq.(\ref{eq:12}) represent the baby loops that 
the universe has a small boundary volume; the last term represents 
the mother loop that the universe has a large boundary volume. 
The distribution function, $\rho(x=L/D^2, D)$ of eq.(\ref{eq:12}), 
satisfies the scaling relation under rescaling 
($D \rightarrow D' = \sqrt{\lambda}D, L \rightarrow L' = \lambda D$): 
\begin{equation}
 \rho(L,D) = \lambda^{-1} \rho(L', D').
\label{eq:13}
\end{equation}

In the four-dimensional model, unfortunately, a similar scaling 
relation about the boundary volume is not yet known. 
We thus assume the distribution function, $\rho(x=V/D^\alpha)$, of 
the volume of the boundary ($V$) within the geodesic distance ($D$) 
in four-dimensional dynamical triangulated manifold. This is a function 
of the scaling variable, $x=V/D^{\alpha}$, with scaling parameter 
($\alpha$) in analogy with the two-dimensional model. 
In Fig.\ref{fig:universe}, we show a schematic picture of our 
boundary analysis.

First, we consider the scaling structures of these three
phases: a crumpled phase, a smooth phase and a branched 
polymer phase\cite{PHASE-DIAGRAM}. 
%
%
%
%
Actually, in the smooth phase we observe that the distribution  
($\rho$) becomes fractal in the sense that the sections of the manifold 
at different distances from a given $4$-simplex look exactly the same 
after a proper rescaling of the boundary volume.
Furthermore, the shape of this scaling function is very similar to that of 
the two-dimensional case\cite{KKMW,TY}.
The best account for this excellent agreement in the four-dimensional case 
can be found in the dominance of a conformal mode and a fractal property.
We have also investigated the boundary volume distribution in the crumpled 
phase and the branched polymer phase.
It seems reasonable to suppose that this new smooth phase has a similar 
fractal structure to that of the two-dimensional fractal surface, and 
that there is a possibility of taking a continuum limit in the phase. 
We have also investigated the boundary volume distribution in both 
the crumpled phase and the branched polymer phase.
In the crumpled phase we find that one mother universe is dominant, 
while in the branched polymer phase there is no evidence for the 
existence of a mother universe.

Next, let us discuss the relation between the scaling parameter 
($\alpha$) and the Hausdorff dimension ($d_H$). 
The expectation value of the boundary three-dimensional volume appearing
at distance $D$ has been introduced in ref.\cite{EHITY}: 
\begin{equation}
 <V^{(3)}>=\frac{1}{N} \int_{v_0}^{\infty} dV \, V \rho(x=V/D^\alpha,D), 
\label{eq:14}
\end{equation}
where $v_0$ denotes the UV cut-off of the boundary volume and $N$ is
the normalization factor. 
If the boundary volume has the scaling property with the universal
distribution ($\rho(x,D)$) and $v_0 \rightarrow 0$, 
\begin{equation}
 <V^{(3)}> \sim D^{\alpha}. 
\label{eq:15}
\end{equation}
Then, we should obtain a finite fractal dimension, 
\begin{equation}
 d_f = \alpha + 1, 
\label{eq:16} 
\end{equation} 
with the fractal dimension $d_f$. 
We measure the volume of the mother boundary as a function of $D$. 
The mother boundary is defined by the boundary having the largest tip 
volume. 
%
\begin{figure}
\centerline{\psfig{file=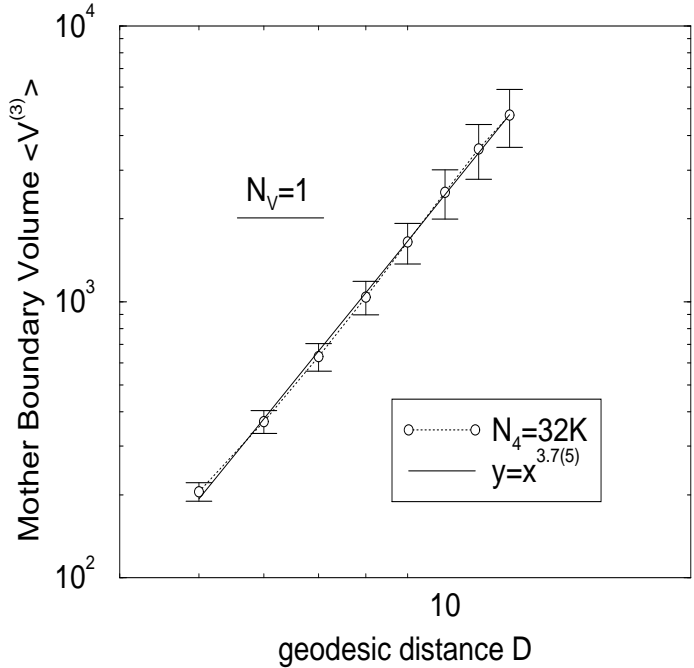,height=9cm,width=9cm}} 
\caption
{
Power fit to the mother boundary three-dimensional volume 
$<V^{(3)}>$ at the critical point ($\kappa_2^{c} = 1.37147(1)$) 
for $N_{V} = 1$ with $N_{4} = 32K$. 
}
\label{fig:mother}
\end{figure}
%
In Fig.\ref{fig:mother}, we plot the mother boundary volume 
$<V^{(3)}>$ with the size of $N_4 = 32K$ at the critical point. 
As a result, the mother boundary volume shows a scaling, and 
we obtain the scaling parameter ($\alpha = 3.7(5)$). 
Then, we can estimate the fractal dimension ($d_f= 4.7(5)$). 
On the other hand, we measured the Hausdorff dimension, 
which results in $d_H = 4.6(2)$. 
Both results are consistent ($d_f \approx d_H$). 
%
%
We thus expect that the boundary volume has a scaling property 
in the sense that the manifold at different distances from 
a given $4$-simplex looks exactly the same after a proper rescaling 
of the boundary volume.
%
%
%
\section{Modified balls-in-boxes model}
%
In this section, we discuss a scenario for the phase structure and the 
phase transition in four-dimensional simplicial quantum gravity coupled to 
matter fields. 
In the pure gravity case, the first order phase transition is described 
by the balls-in-boxes model\cite{BBPT,BBJ,BB}. 
This model is considered to be a simple mean field model about 
simplicial quantum gravity. 
It describes a fixed number $N$ balls distributed into a variable $M$ of
boxes. 
The partition function is given by 
\begin{equation}
 Z_{M,N} = \sum_{q_1,...,q_M} p(q_1) \cdots p(q_M) 
\delta_{q_1+\cdots+q_M,N}, 
\label{eq:17} 
\end{equation} 
where $q_{i}$ is the number of balls in box $i$, and $p(q)$ denotes a
weight which depends on only the number of balls (q) in the box.
The original balls-in-boxes model has the following correspondence: 
\begin{center} 
Vertex $\leftrightarrow$ Box and Simplex $\leftrightarrow$ Ball. 
\end{center} 
Therefore, $q_{i}$ represents the vertex order. 
The original model has been analyzed in the case of power like 
weights: $p(q) = q^{-\beta}$ (when $\beta$ is a parameter for a 
weight). The system has two distinct phases: a crumpled phase 
and an elongated phase, and it has a discontinuous phase transition 
for $\beta > 2$. 
It describes the phase structure about four-dimensional pure simplicial 
quantum gravity. 
However, this model describes the continuous phase transition at the 
$\beta < 2$ \cite{BB}. 
We have noticed this fact, which gives one of the motivations for 
the modified model of simplicial quantum gravity. 
In the model for simplicial quantum gravity, we can consider
that parameter $\beta$ is related to the measure term. 
If we consider the conformal matter fields (as the conformal charge), 
parameter $\beta$ has the same origin as the conformal 
anomaly, because they are related to the measure term. 
The gauge field gives rise to a modified measure 
factor\cite{BILKE2}, 
\begin{equation}
  M_{t} = \prod_{t} o(t)^{\beta}.
\label{eq:18}
\end{equation}
This model shows three distinct phases: a crumpled phase, 
a crinkled phase (correspond to a smooth phase) and 
a branched polymer phase. 
It is consistent with our results. 
The natural correspondence between our phase structure and that in 
Ref.\cite{BILKE2} is as follows: the transition between the crumpled 
phase and the branched polymer phase is discontinuous (probably, 
first-order phase transition), the transition between the crumpled 
phase and the crinkled phase is continuous, and the transition 
between the crinkled phase and the branched polymer phase is 
obscure. 

Thus, in order to investigate the phase structure of adding matter 
fields, we exchange the relation of ``vertices-simplices'' into 
that of ``triangle-simplices'': 
\begin{center}
Triangle $\leftrightarrow$ Box and Simplex $\leftrightarrow$ Ball. 
\end{center}
We thus introduce the triangle order ($o(t)$) instead of 
the vertex order ($q$).
This modification causes the partition function to give rise 
a constant shift for the constraint: 
\begin{equation} 
\sum_{i:vertex} q_{i} = 5N_{4},
\end{equation}
\begin{equation} 
\sum_{t:triangle} o(t) = 10N_{4}.
\end{equation}
We analyzed the modified balls-in-boxes model with using the same
method of the both original models. 

We thus considered the case of simplicial quantum gravity coupled 
to matter fields. 
The partition function is 
\begin{eqnarray}
Z_{N}(\kappa, \beta, N_{matter}) 
&=& \sum_{m}^{M_{max}} e^{\kappa m} \nonumber \\
& & \sum_{q_1,...q_m} p(q_1) \cdots p(q_m) \delta_{q_1+\cdots+q_m,N} 
(Z_{mattar}(m))^{N_{mattar}}.
\label{eq:20}
\end{eqnarray}
Now we discuss the following two cases: (a) the case of coupling to 
$N_B$ copies of the boson fields and (b) the case of coupling to $N_V$ 
copies of the vector fields. 
(a) In the first case, the matter action is 
\begin{equation}
 S_{boson} = \sum_{(ij)}(\phi_i - \phi_j)^2. 
\end{equation}
We discuss the effect from the matter. 
From the perturbation, the Gaussian boson matter gives
\begin{equation}
 Z_{boson} = \sum_{triangles} Constant.
\end{equation}
We can estimate the effect to the weight of one-boxes: 
\begin{equation}
 p(q) \rightarrow C \cdot p(q). 
\end{equation}
The Gaussian bosons give the effect of multiplication by a constant.
From a mean fields analysis, the phase diagram will be changed.
However, it is known that the phase diagram is not changed by 
only a few Gaussian boson fields, according to numerical 
analysis\cite{JBJK}. 
We consider that the multiplication by a constant is not enough 
to change the phase diagram for ``real'' simplicial quantum gravity, 
because of the too small effect of multiplication by a constant. 
(b) In the second case, we consider adding the Gaussian $U(1)$ gauge 
matter.
If we use the plaquette action for the gauge fields, the partition
function is written as 
\begin{equation}
 Z_{gauge} = \sum_{triangle} f(o(t)),
\end{equation}
where $f$ is a function of the triangle order ($o(t)$). 
From the perturbation about the gauge interaction, 
$f(o(t)) = C_{0} + C_{1}o(t)$. 
%
%
Then, the partition function is replaced, as follows: 
\begin{equation} 
 Z_{gauge} \sim \sum_{t} o(t).
\end{equation}
This is just the modified measure.
We estimate the effect from the gauge matter fields as 
\begin{equation}
 \beta \rightarrow (\beta - 1). 
\end{equation}
That is to say, the Gaussian gauge matter fields lower $\beta$.
%
\section{Summary and Discussion}
%
We have investigated the phase structure and the phase transition 
with a model of four-dimensional simplicial quantum gravity coupled 
to $U(1)$ gauge fields. 

The results of our study are summarized by the schematic phase diagram in 
Fig.\ref{fig:phasediagram}. We checked this phase diagram in the case of
$N_{V}=1,2,3$ at a volume of $N_4=16K$. 
We found three phases in this parameter space: a crumpled
phase ($\gamma_{st} \rightarrow \infty$, $d_f \rightarrow \infty$), a
smooth phase ($\gamma_{st} < 0$, $2< d_f \sim 4$, $N_0/N_4 < 0.25$) and a
branched polymer phase ($\gamma_{st} > 0$, $d_f = 2$, $N_0/N_4 = 0.25$).
The thin line denotes a discontinuous phase-transition line which is 
known in pure gravity; moreover, the thick line denotes a smooth 
phase-transition line. 
As contrasted with the phase diagram of pure gravity, the phase 
diagram means richer structures. 
In the crumpled phase one can find singular sub-simplices, for example, 
vertex order concentration and link order concentration. 
The smooth phase is defined as a region between the critical point 
($\kappa_2^c$) and the obscure phase transition point ($\kappa_2^o$).
We observe the negative string susceptibility in this region. 
Then, with $N_{4}=16K$ ($\kappa_{2}=1.7$) and $N_{V}=1$, 
we obtained $\gamma_{st}=-0.38(5)$, $d_{f}=2.8(5)$ and 
a good scaling relation of the boundary volume distributions. 
We consider that the scaling structure of this smooth phase is 
similar to that of a two-dimensional random (fractal) surface. 
This smooth phase is slowly broken to a branched polymer phase 
that has no mother structure at the obscure phase transition 
($\kappa_2^o$).
We obtained an obscure-transition line (a broken line in 
Fig.\ref{fig:phasediagram}); therefore, we suggest that the 
obscure-transition corresponds to the $c=1$ barrier 
in two-dimensional quantum gravity.

As for the phase transition at the critical point ($\kappa_2=\kappa_2^c$),
we showed the finite size scaling at the critical point and the histogram
of the node. 
We calculated some finite size scaling exponents, and showed that 
the value of these exponents is smaller than 1. 
It has been discussed \cite{BAKKER} that the value of 1 is expected to be 
a first order phase transition for the results of pure gravity.
However, in the case of adding one gauge matter field ($N_V=1$), 
the numerical results show that the phase transition is smooth, 
in the contrast to pure gravity, and then that a double peak structure 
of the node histogram disappears. 
%
%
%
%

Furthermore, in order to investigate the property of quantum geometry
and to discuss the universality of the manifold, we considered 
the scaling property of the boundary volume. 
In the smooth phase with $N_{4}=16K$ ($\kappa_{2}=1.7$) and $N_{V}=1$, 
we obtained $d_{f}=2.8(5)$ and a good scaling relation of the boundary
volume distributions. 
The scaling structure of this smooth phase is similar to that of a
two-dimensional random (fractal) surface.
This suggests the existence of a new smooth phase in four-dimensional 
simplicial gravity. 
The other two phases have a similar scaling property to that of 
pure gravity. 
We have shown the scaling property of the mother boundary, 
where the scaling parameter is consistent with the Hausdorff dimension. 
Then, the boundary volume has a scaling property 
with the scaling variable ($x=V/D^{d_{f}-1}$). 
We expect that the boundaries have a fractal structure and universality
of the scaling relations. 
%
%
%
%
We also discussed the modification of the balls-in-boxes model. 
The role of a vertex is exchanged into a triangle. 
This clarifies the relation between the measure factor of 
the numerical model and that of the analytical mean field model.

We expect the existence of genuine four-dimensional quantum gravity 
on the critical point ($\kappa_{2}^{c}$) with the vector fields.
Our recent investigations will give further evidence 
for the existence of an ultraviolet fixed point of 
the quantum field theory of gravity.
\section{Acknowledgments}
We would like to thank H.Kawai, N.Ishibashi, K.Hamada and F.Sugino 
for fruitful discussions. We are also grateful to members of 
the KEK-IPNS theory group. The numerical calculations were performed 
using the originally designed cluster computer for the study about 
quantum gravity and strings; CCGS-01 ATROPOS(Tokai University), 
CCGS-02 EST and CCGS-03 LACHESIS(KEK).

\end{document}